# Biases and Limits of AlphaFold Predictions for Non-Globular Proteins

*Maria Cristina Aspromonte[1], Alexander M. Monzon[1], Miguel A. Andrade-Navarro[2], Oriol Bárcenas[3], Viktor Bartosik[4], Pau Bernadó[5], Juan Cortés[6], Zsuzsanna Dosztanyi[7], Gabor Erdos[8], Isabella C. Felli[9], Helena Gradišar[10], San Hadži[11], Tamás Hegedűs[12], Khalil Joron[13], Pavel Kaderavek[4], Eitan Lerner[13], Dirk Linke[14], Pablo Mier[15], Ana Pantelić[16], Roberta Pierattelli[9], Salvador Ventura[3], Marija Vidović[16], Metaxia Vlassi[17], Lukas Zidek[4] and Silvio C.E. Tosatto[1,*]*

*[1] - Department of Biomedical Sciences, University of Padova, Italy*
*[2] - Institute of Organismic and Molecular Evolution, Faculty of Biology, Johannes Gutenberg University of Mainz, Germany*
*[3] - Institut de Biotecnologia i Biomedicina and Departament de Bioquímica i Biologia Molecular, Universitat Autònoma de Barcelona and Institute of Advanced Chemistry of Catalonia (IQAC), CSIC, Spain*
*[4] - National Centre for Biomolecular Research, Faculty of Science and Central European Institute of Technology, Masaryk University, Czech Republic*
*[5] - Centre de Biologie Structurale, Université de Montpellier, INSERM, CNRS*
*[6] - Université de Toulouse, CNRS, LAAS, Toulouse, France*
*[7] - CRB, Univeristè de Montpellier, Montpellier, France*
*[8] - Department of Biochemistry, Eötvös Loránd University, Budapest, Hungary*
*[9] - Center for Magnetic Resonance (CERM) and Department of Chemistry Ugo Schiff (DICUS), University of Florence, Italy*
*[10] - Department of Synthetic Biology and Immunology, National Institute of Chemistry, Ljubljana, Slovenia*
*[11] - Department of Physical Chemistry, Faculty of Chemistry and Chemical Technology, University of Ljubljana, Slovenia*
*[12] - Department of Prosthodontics, Semmelweis University, Budapest, Hungary*
*[13] - The Department of Biological Chemistry, The Alexander Silberman Institute for Life Sciences, Faculty of Sciences, The Edmond J. Safra Campus, The Hebrew University of Jerusalem, Israel*
*[14] - Department of Biosciences, University of Oslo, Oslo, Norway*
*[15] - Andalusian Centre for Developmental Biology (CABD, UPO-CSIC-JA), Faculty of Experimental Sciences, Universidad Pablo de Olavide (Seville), Spain*
*[16] - Department of Microbiology and Plant Biology, Institute of Molecular Genetics and Genetic Engineering (IMGGE), University of Belgrade, Serbia*
*[17] - Institute of Biosciences & Applications, National Center for Scientific Research "DEMOKRITOS", Athens, Greece*
*[*] - To whom correspondence should be addressed at: silvio.tosatto@unipd.it*

**AlphaFold has transformed protein structure prediction, making structural models accessible across the life sciences. We explore experimentally validated cases of non-globular proteins to uncover where predictions diverge from biologically relevant structural states and how molecular context shapes these discrepancies, guiding a more informed interpretation of AlphaFold models.**

## Beyond a single structure: AlphaFold and NGPs

In recent years, enormous progress has been made in protein structure prediction with the development of AlphaFold (AF)[1] and building on its concepts and architecture. AlphaFold introduced a novel neural network architecture able to capture long-range relationships between amino acids by exploiting the evolutionary information encoded in multiple sequence alignments. AlphaFold2 (AF2)[2] substantially advanced this framework through an end-to-end neural network architecture. In May 2024, AlphaFold3 (AF3)[3] was released, introducing a substantially updated architecture based on diffusion modelling, a generative approach that progressively refines structural predictions. This enables AF3 to predict the joint structure of complexes containing proteins, nucleic acids, small molecules, ions and modified residues. The accuracy of these predictions, typically assessed by how closely they match structures in Protein Data Bank (PDB)[4], can vary substantially and does not account for the inherent flexibility and dynamic nature of proteins. Indeed, a substantial fraction of the human proteome with low-confidence AlphaFold predictions are likely to contain regions that lack stable 3D structures but assume rapidly interconverting, dynamic ensembles of conformations. These Non-Globular Proteins (NGPs) encompass a broad and heterogeneous group, including Intrinsically Disordered Proteins (IDPs), Repeat Proteins, and Low Complexity Regions, among others[5,6]. In this regard, the possibility of leveraging AlphaFold to predict protein disorder, for example, has been systematically explored within the framework of the Critical Assessment of Intrinsic Disorder (CAID) experiments, which benchmark computational methods for disorder prediction with experimental data deposited in the DisProt database[7]. Results from CAID3[8], and the systematic comparisons of AF2[2] and AF3[3] on experimentally annotated disordered regions, have shown that AlphaFold-derived metrics, such as pLDDT (AlphaFold-pLDDT) and RSA (AlphaFold-RSA), can provide indirect signals of disorder but do not fully represent the conformational heterogeneity observed experimentally.

Nevertheless, it is difficult to remain unimpressed by AlphaFold's progress. AlphaFold has become a go-to tool for both experts and non-experts to generate structural models of proteins. While experts can fully interpret the limitations of the predictions based on the complex "background" outputs, non-expert users might unintentionally overinterpret these structural snapshots without considering their inherent uncertainties. In discussion groups at conferences and workshops of the COST Action ML4NGP (Machine learning for non-globular proteins; URL: https://ml4ngp.eu/), this was recognized as a potential risk that could compromise the quality of publications, grant applications, and other scientific output that the experts review. Consequently, a two-day workshop held in Barcelona in March 2025 was used as a forum to collect examples of incorrect AlphaFold models for NGPs to caution scientists against blindly accepting AlphaFold results without critical evaluation. Although the examples collected in this work span diverse proteins, biological processes and structural scales, they reveal a common theme: many apparent discrepancies between AlphaFold models and experimental observations do not arise from inaccurate geometry prediction *per se*, but from limitations in capturing how biological context shapes protein conformation. As an example, α-Synuclein, a protein central to Parkinson's disease, captures much of this complexity: its structural landscape changes dramatically with its biological context (Supplementary Figure 1). In solution, it behaves as an intrinsically disordered monomer, while membrane association promotes the formation of α-helical segments and self-association gives rise to oligomeric and ultimately fibrillar assemblies.

In many NGPs, the experimentally observed structural state depends on information that extends beyond the primary amino acid sequence, including molecular interactions, environmental conditions, post-translational modifications (PTMs) and transitions between alternative conformational states. Although AF3 can explicitly incorporate additional molecular components and selected chemical modifications, this does not necessarily translate into an accurate representation of context-dependent conformational states, particularly for NGPs. Here, we aim to provide a framework for users and developers to interpret such predictions in a wider biological context and in conjunction with complementary experimental evidence.

## Evaluating AlphaFold predictions against experimental structural data for NGPs

To examine these limitations, we analyzed a collection of 27 experimentally validated cases contributed by leading experts in the field of NGPs (see Supplementary Table S1, Supplementary Figures). These were grouped into four major categories, corresponding to distinct structural and functional levels at which AlphaFold diverges from experimental observations: (i) Interaction and assembly (ii) Biological and environmental context (iii) Global architecture and topology (iv) Membrane topology. Together, these categories move from the correct assembly of protein components and the representation of their biologically relevant states, to their global organization, including the emerging challenge of predicting non-natural, designed protein architectures, and finally to their compatibility with physical constraints such as the membrane environment. To illustrate the four major classes of discrepancies, we selected one representative example for each category that exemplifies a common challenge in the biological interpretation of AlphaFold predictions (Fig. 1).

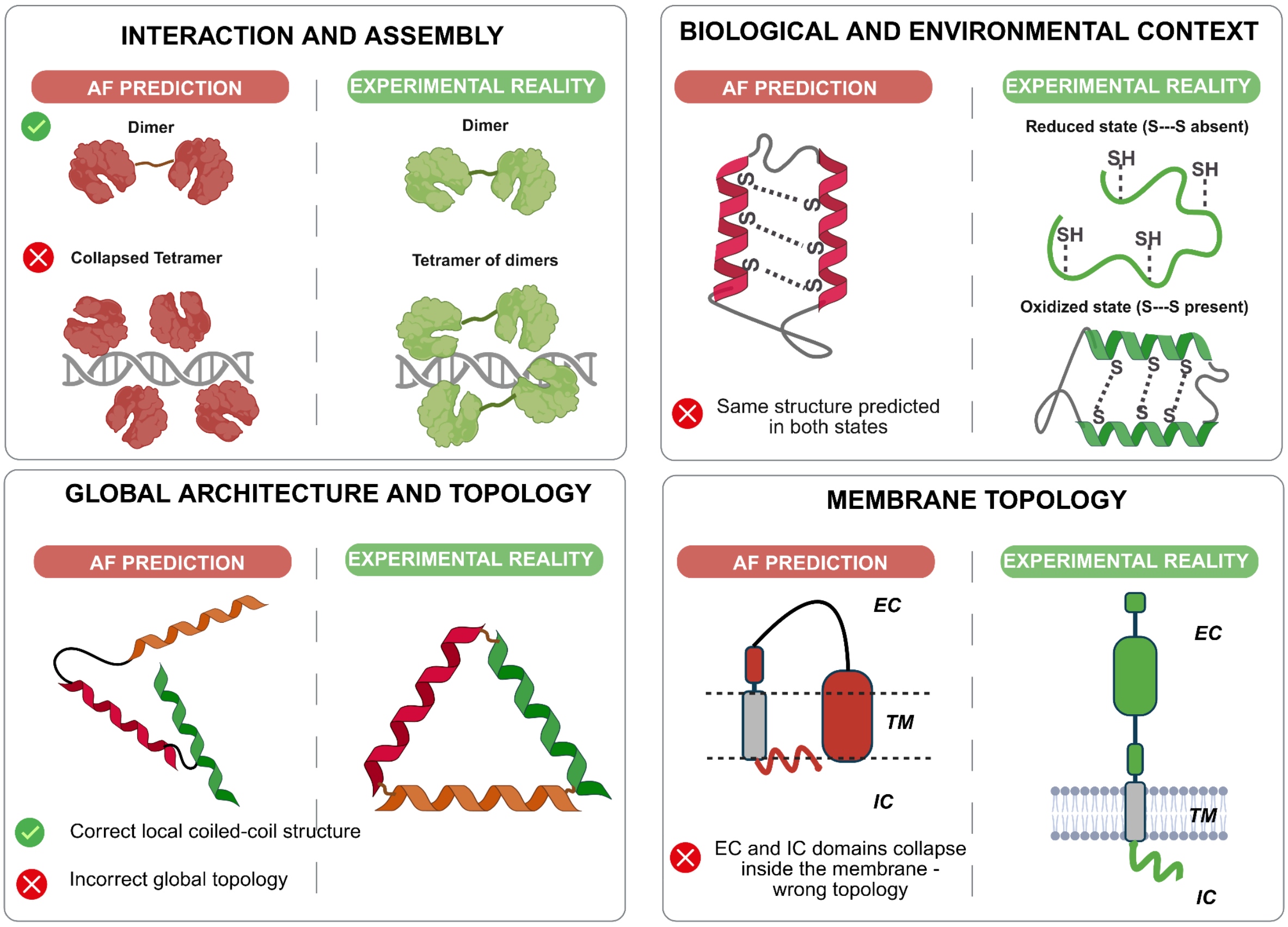

**Figure 1** Representative examples illustrating how AlphaFold predictions can misrepresent NGPs. The four examples illustrate challenges related to interaction and assembly (p53 tetramerization), biological context and environmental context (redox-dependent conformational states of TRIAP1), global architecture and topology (Coiled-Coil Protein Origami, CCPO), and membrane topology (Intimin). Abbreviations: AF, AlphaFold; CC, Coiled-coil; EC, Extracellular; IC, Intracellular; TM, Transmembrane.

# Interaction and assembly

Many NGPs do not adopt a single biologically relevant conformation. The thermodynamics of interactions involving NGPs are complex. Their structural state depends on the molecular context, including interactions with binding partners or assembly into higher-order complexes. This structural plasticity is particularly important for IDPs and IDRs, which play central roles in molecular recognition and assembly of regulatory complexes. IDRs can be highly multivalent, with several Short Linear Motifs (SLiMs) or repeats orchestrating higher-order complex formation[9].

At the interaction level, the interpretation of AlphaFold predictions becomes more challenging for proteins participating in complex quaternary assemblies or undergoing conformational changes upon binding. Proteins that change structure upon interaction or assemble into dynamic complexes often defy this assumption. The tetrameric tumor suppressor p53 represents a major challenge for 3D structure prediction because of the complex spatial organization of its dimeric and tetrameric assemblies (Supplementary Fig.S2). p53 first forms dimers through its tetramerization domain (TET), and two dimers further associate through the same domain to form the functional tetramer. The tetramer then cooperatively binds target duplex DNA in a sequence-specific manner through its DNA-binding domains (DBDs). The dimeric architecture is predicted remarkably well by AlphaFold, closely reproducing the experimentally determined arrangement of the TET domains. However, prediction of the tetrameric complex highlights a limitation in modelling the oligomeric state: while the individual dimeric units are well represented, their relative arrangement in the higher-order tetramer differs from the experimentally determined assembly.

# Biological and environmental context

Many IDPs adopt a broad conformational ensemble of interconverting states, disorder-to-order transitions and vice versa only upon interaction with their physiological partners, a process known as folding-upon-binding[10] or fuzzy complexes[11]. Substantial effort has been devoted to characterizing the kinetics and thermodynamics of these binding events, combining NMR, optical spectroscopy, calorimetry, and molecular dynamics simulations to understand how flexibility underlies specificity[12]. However, interaction-dependent structural behaviour is not limited to local folding events. Consequently, many NGPs do not possess a single biologically relevant structure but rather populate a set of states whose distribution depends on their interaction partners and molecular environment. Biological context and molecular state encompass cases in which the same protein sequence can adopt different structural states depending on its biological context. These include redox conditions, PTMs, proteolytic processing, ligand binding, and other cellular factors. Because AF2 predicts structures from sequence alone, it cannot determine which of these biological states is present, and therefore represents only one possible structural state. AF3 extends the modelling framework to post-translationally modified residues as input, among others. Nevertheless, many biologically relevant molecular states remain undefined by sequence alone and cannot be explicitly modelled, including redox-dependent conformational changes, proteolytic processing and other context-dependent structural transitions. We therefore examined examples in which experimentally observed structural states are controlled by biological factors not fully specified by the amino acid sequence, including PTMs, proteolytic processing and redox conditions. Redox-dependent proteins provide a representative use case. TRIAP1 is a disulfide-rich protein, whose conformation depends on the redox state of its cysteine residues. The oxidized protein adopts a predominantly α-helical structure, whereas the reduced form resembles a molten globule. However, AlphaFold predictions remain essentially unchanged even when cysteines are replaced by serines to mimic the reduced state (Supplementary Fig. S7).

# Global architecture and topology

Global architecture and topology encompasses challenges in reconstructing the overall organization of a protein chain, including the correct connectivity and spatial arrangement of structural elements, as well as the organization and periodicity of repeat-containing regions. These aspects are particularly relevant for proteins with flexible multidomain architectures and unusual or highly constrained topologies, and are becoming increasingly important as *de novo* protein design expands the structural space towards complex, non-natural architectures[13].

Here, we report a particularly illustrative case involving a *de novo* Coiled-Coil Protein Origami (CCPO) (Supplementary Fig. S16). In CCPO, a single polypeptide chain is designed to self-assemble into a predefined polyhedral structure through the specific pairing of coiled-coil segments. Consequently, the final architecture depends not only on the correct local formation of individual coiled coils, but also on the precise connectivity and topological arrangement of the entire chain.

AlphaFold can reproduce several local structural features of the coiled-coil segments, while the predicted global arrangement differs from the experimentally determined topology. This example highlights an important challenge for designed proteins: accurate prediction of local structural elements does not necessarily guarantee correct reconstruction of the intended global topology. As increasingly complex non-natural protein architectures are developed, reliable prediction of their overall topology becomes particularly important for the evaluation and validation of protein designs.

# Membrane topology

This last category represents a specific challenge for structure prediction because their correct biological organization depends not only on the local accuracy of individual domains but also on their topology relative to the membrane. The membrane defines the orientation and spatial separation of extracellular, transmembrane, periplasmic, and cytoplasmic regions, and therefore provides essential information for determining the overall topology of the protein. Since AlphaFold does not explicitly model the membrane environment, predicted domains can adopt locally plausible structures while being arranged in positions or orientations that are incompatible with the membrane[14].

Type Ve autotransporters from different Gram-negative bacteria are transmembrane beta-barrel proteins with an extended, extracellular part composed of repeating Ig domain units and a C-terminal lectin domain that mediates binding to host cells in infection processes, and a periplasmic domain known to mediate dimerization and sometimes, peptidoglycan binding. Structures of individual domains and domain groups are available, but no structure has been solved experimentally for a complete Type Ve autotransporter yet. Both AF2 and AF3 predict multi-domain arrangements where extracellular domains would cross the membrane plane, and where the periplasmic domain is not correctly predicted. When enforcing dimer predictions with AF3, the topology of extracellular domains relative to the membrane is again wrong, and the predicted dimerization interface is in conflict with the membrane plane. Additionally, periplasmic coiled-coil segments are not predicted as dimers and also have wrong topological features (Supplementary Fig.S21).

# Final considerations

AlphaFold has revolutionized biology and biomedicine by providing large-scale access to protein structure predictions of unprecedented accuracy. At the same time, our collection of experimentally validated examples highlights an important distinction between predicting a protein structure and interpreting its biological meaning. These limitations are particularly relevant for NGPs, whose structural organization may depend on conformational heterogeneity, molecular interactions, global topology, membrane environment, or specific biological and biochemical states. These factors define structural states and ensembles that cannot always be represented by a single static model.

The four categories proposed here provide a practical framework for interpreting AlphaFold predictions. They encourage users to consider whether a predicted model may depend on interaction partners, membrane topology or domain organization, local structural heterogeneity, or a specific biological or biochemical state before drawing functional conclusions. In many cases, complementary experimental evidence, including NMR, SAXS, cryo-EM, cross-linking or circular dichroism, remains essential for selecting the biologically relevant structural interpretation[15].

The examples collected here also allow us to consider which challenges persist across different generations of AlphaFold. α-Synuclein further illustrates this limitation: AF3 predictions vary with template use, yet even a fibril template produces a fibril-like structure lacking the characteristic β-sheet architecture.

AF3 introduces important methodological advances, including improved modelling of biomolecular complexes and the possibility of specifying selected post-translational modifications as input. Nevertheless, comparison across the examples presented here shows that many of the major challenges remain shared between both generations of the method. In particular, proteins that undergo conformational transitions from intrinsically disordered to ordered states upon binding to molecular partners or changes in their molecular environment remain difficult to predict consistently.

While we are hopeful that the future of protein structure prediction will progress towards more accurate results for NGPs considering the four points discussed above, there are some steps in the way that will require some thinking. Mainly, we need to go beyond the concept of one protein sequence corresponding to one structure. We need structural predictions that consider conformational variability. But how to represent such data? How to store predictions of variable structures? Which type of experimental and computational data can be used to train methods to produce such predictions? How can we include in these predictions environmental conditions (*e.g.*, pH, ionic strength, redox state, etc.) and other elements influencing structures such as the lipidic membrane or other macromolecules in a standardized way?

Without any doubt, while these problems are addressed, AlphaFold will be extensively used (most successfully for the most part). With this paper, we pointed to salient problems that pertain to our expertise in NGPs, informing everyday users of where the most important sources of errors in AlphaFold predictions are, but also showing a way forward towards a better future of protein structure prediction.

# Acknowledgements

The authors are grateful to Lorenzo Bracaglia, Mahta Mehdiabadi, Damiano Piovesan and Marco Schiavina for insightful discussions.
This work was supported by: COST Action ML4NGP (CA21160), supported by COST (European Cooperation in Science and Technology); European Union under grant agreement no. 101160233 (HORIZON-Twinning project IDP2Biomed) and under grant agreement no. 101182949 (HORIZON-MSCA-SE project IDPfun2); funding from EKA:2025/041-P108-1; MICIU/AEI/10.13039/501100011033 grant No. BG23/00060, Slovenian Research Agency grants P1-0201 and J1-50026, Czech Science Foundation grant No. GA25-16037S and the French National Research Agency (ANR) under grant ANR-22-CE45-0003.
Views and opinions expressed are however those of the author(s) only and do not necessarily reflect those of the European Union or the European Research Executive Agency (REA). Neither the European Union nor the granting authority can be held responsible for them.

## Ethics declarations

All other authors declare no competing interests.

## Supplementary information

Supplementary Tables 1, Supplementary Figs. 1– 24.